\documentclass[aps,pra,twocolumn,superscriptaddress,groupedaddress]{revtex4}  
\usepackage{graphicx}  
\usepackage{dcolumn}   
\usepackage{bm}        
\usepackage{amssymb}   
\usepackage{upgreek}
\usepackage{subfigure}
\usepackage{amsmath}
\usepackage{braket}
\usepackage{color}

\begin{document}


\title{Phonon Interferometry for Measuring Quantum Decoherence}

\author{M. J. Weaver}
\email{mweaver@physics.ucsb.edu}
\affiliation{Department of Physics, University of California, Santa Barbara, California 93106, USA}
\author{D. Newsom}
\affiliation{Department of Physics, University of California, Santa Barbara, California 93106, USA}
\author{F. Luna}
\affiliation{Department of Physics, University of California, Santa Barbara, California 93106, USA}
\author{W. L\"{o}ffler}
\affiliation{Huygens-Kamerlingh Onnes Laboratorium, Universiteit Leiden, 2333 CA Leiden, The Netherlands}
\author{D. Bouwmeester}
\affiliation{Department of Physics, University of California, Santa Barbara, California 93106, USA}
\affiliation{Huygens-Kamerlingh Onnes Laboratorium, Universiteit Leiden, 2333 CA Leiden, The Netherlands}

\date{\today}

\begin{abstract}
Experimental observation of the decoherence of macroscopic objects is of fundamental importance to the study of quantum collapse models and the quantum to classical transition. Optomechanics is a promising field for the study of such models because of its fine control and readout of mechanical motion. Nevertheless, it is challenging to monitor a mechanical superposition state for long enough to investigate this transition. We present a scheme for entangling two mechanical resonators in spatial superposition states such that all quantum information is stored in the mechanical resonators. The scheme is general and applies to any optomechanical system with multiple mechanical modes. By analytic and numeric modeling, we show that the scheme is resilient to experimental imperfections such as incomplete pre-cooling, faulty postselection and inefficient optomechanical coupling. This proposed procedure overcomes limitations of previously proposed schemes that have so far hindered the study of macroscopic quantum dynamics.
\end{abstract}

\maketitle 

\section{Introduction}

The transition between quantum and classical regimes, particularly in massive systems is still largely unexplored. The fields of opto- and electromechanics have emerged as effective tools for controlling and measuring the quantum motion of mechanical resonators \cite{Aspelmeyer2014}. In recent years macroscopic mechanical resonators have been developed with exceptionally high quality factors \cite{Goryachev2012,Tsaturyan2016,Yuan2015}. At the same time devices with a single photon strong cooperativity \cite{Reinhardt2016,Norte2016,Leijssen2017} are enabling manipulation of optomechanical systems at the single quantum level \cite{OConnell2010,Lecocq2015,Riedinger2015}. Large mechanical resonators are proposed to undergo a number of unconventional decoherence mechanisms \cite{Bassi2013,Bassi2003,Diasi1989,Penrose1996}. One promising technique for testing decoherence is to produce a spatial superposition state of one of these resonators, but this requires a controlling interaction with some other quantum system. We investigate a method for entangling two mechanical resonances and harnessing the advantageous capabilities of each resonator to study decoherence.

There are many proposed methods of producing a superposition state in an opto- or electromechanical system, all of which require the introduction of some nonlinearity. Examples of this include electromechanical systems coupled to a superconducting qubit \cite{OConnell2010,Lecocq2015,Chu2017} and optomechanical systems interacting with a single photon sent through a beam splitter \cite{Marshall2003}. However, the latter scheme is unfeasible with almost all current optomechanical systems, because it requires single photon strong coupling \cite{Marshall2003}. This requirement can be circumvented by postselection \cite{Pepper2012} or displacement \cite{Sekatski2014}, but these experiments are limited by the need for long storage of photons, which is lossy, and the requirement that cavity photons predominantly couple to a single mechanical mode. Here we propose a method to eliminate these constraints by entangling two mechanical modes optomechanically to avoid the losses and decoherence in optical and electrical systems.

\begin{figure}
\includegraphics[width=8cm]{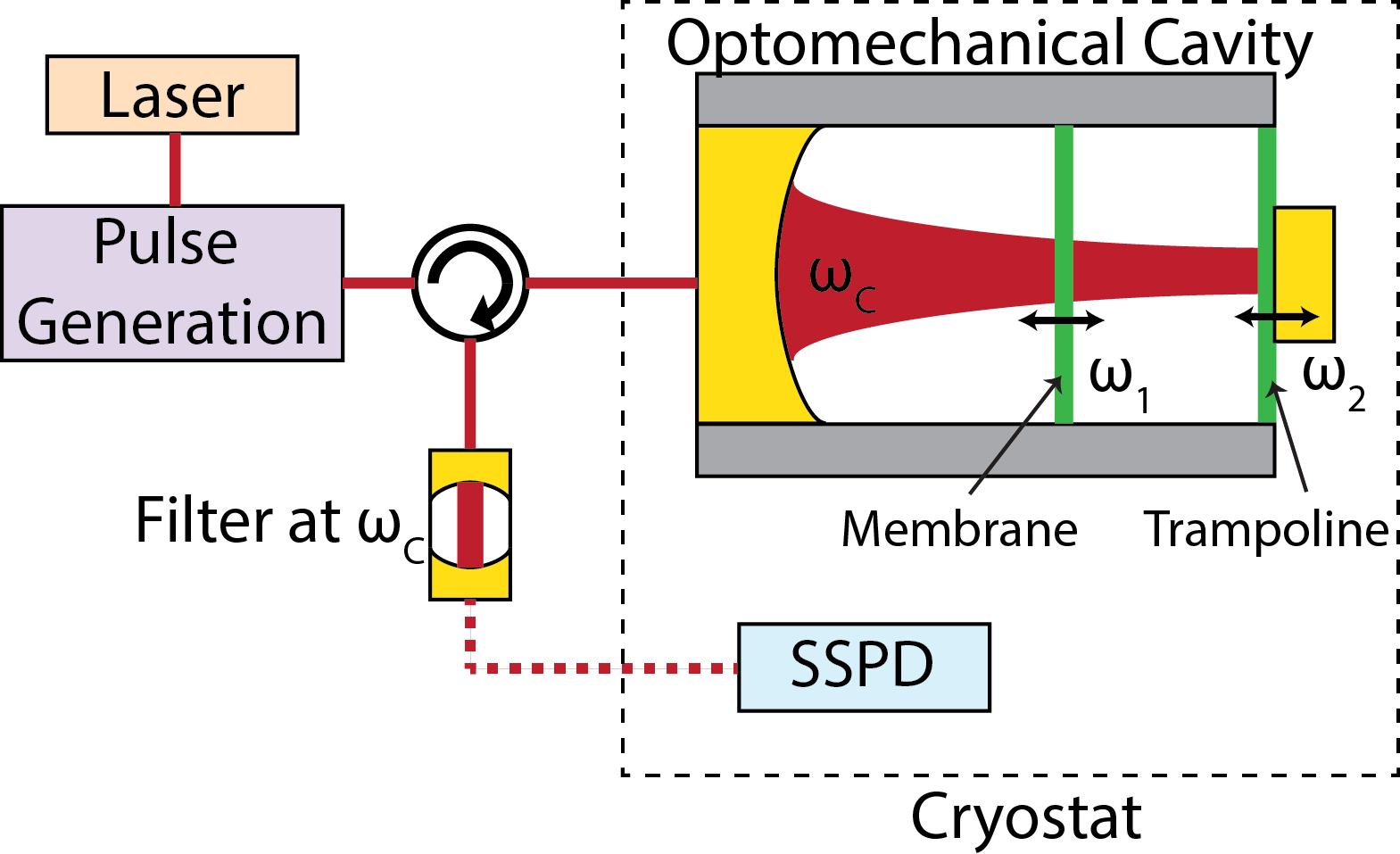}
\caption{Proposed experimental setup. Two mechanical resonators are optomechanically coupled to an optical cavity. Here we show a membrane and a trampoline resonator with a mirror, but the procedure could be used for any two mechanical resonators coupled via an optical cavity field. A continuous wave laser is sent to an optical pulse generation setup, which produces pulses of varying frequency, duration, and intensity. The light enters the optomechanical cavity, and subsequently the reflected light is filtered to remove the control pulses. The filtered signal contains the single photons used for heralding and readout, which are measured with a superconducting single photon detector (SSPD).}
\label{fig:setup}
\end{figure}

Methods to generate optomechanical entanglement between multiple mechanical devices have been investigated extensively \cite{Vitali2007,Hartmann2008,B??rkje2011,Woolley2014,Li2015,Li2017,Zhang2017}. To generate a superposition, an interaction with two mechanical resonators is required \cite{Akram2013,Flayac2014}. So far demonstrations of entanglement in optomechanical systems have used elements with similar structure and frequency \cite{Blatt2008,Lee2011,Riedinger2017}. Flayac and Savona suggested that single photon projection measurements could generate an entangled superposition state between two resonators of similar frequency \cite{Flayac2014}. We propose a scheme which entangles resonators of different frequencies, so that it is easy to manipulate one resonator and to use the other (possibly more massive) resonator for tests of quantum mechanics.

\section{Experimental Scheme}

We consider an optomechanical system with one optical cavity and two mechanical resonators: an interaction resonator (resonator 1) and a quantum test mass resonator (resonator 2). The Hamiltonian for the system is the standard optomechanics Hamiltonian for multiple resonators \cite{Aspelmeyer2014}:

\begin{equation}
\hat{H}_0= \hbar\omega_c \hat{a}^\dagger\hat{a} + \sum_{j=1,2} \hbar\omega_{j}\hat{b}_j^\dagger\hat{b}_j + \hbar g_j\hat{a}^\dagger\hat{a}(\hat{b}_j^\dagger + \hat{b}_j)
\end{equation}
$\omega_c$, $\hat{a}$, $\omega_j$, $\hat{b}_j$ are the frequencies and bosonic ladder operators of the cavity and resonator $j$ respectively. $g_j$ are the single photon optomechanical coupling rates. The system is sideband resolved, with $\omega_j \gg \kappa$, the optical cavity linewidth. In Figure \ref{fig:setup} the optomechanical setup is shown. A laser is modulated to generate control pulses, for instance by a series of acousto-optic modulators (AOMs). The pulses are sent into the cavity, and are filtered out of the light exiting the cavity so that only the remaining resonant light is incident on a single photon detector.

\begin{figure}
\includegraphics[width=8cm]{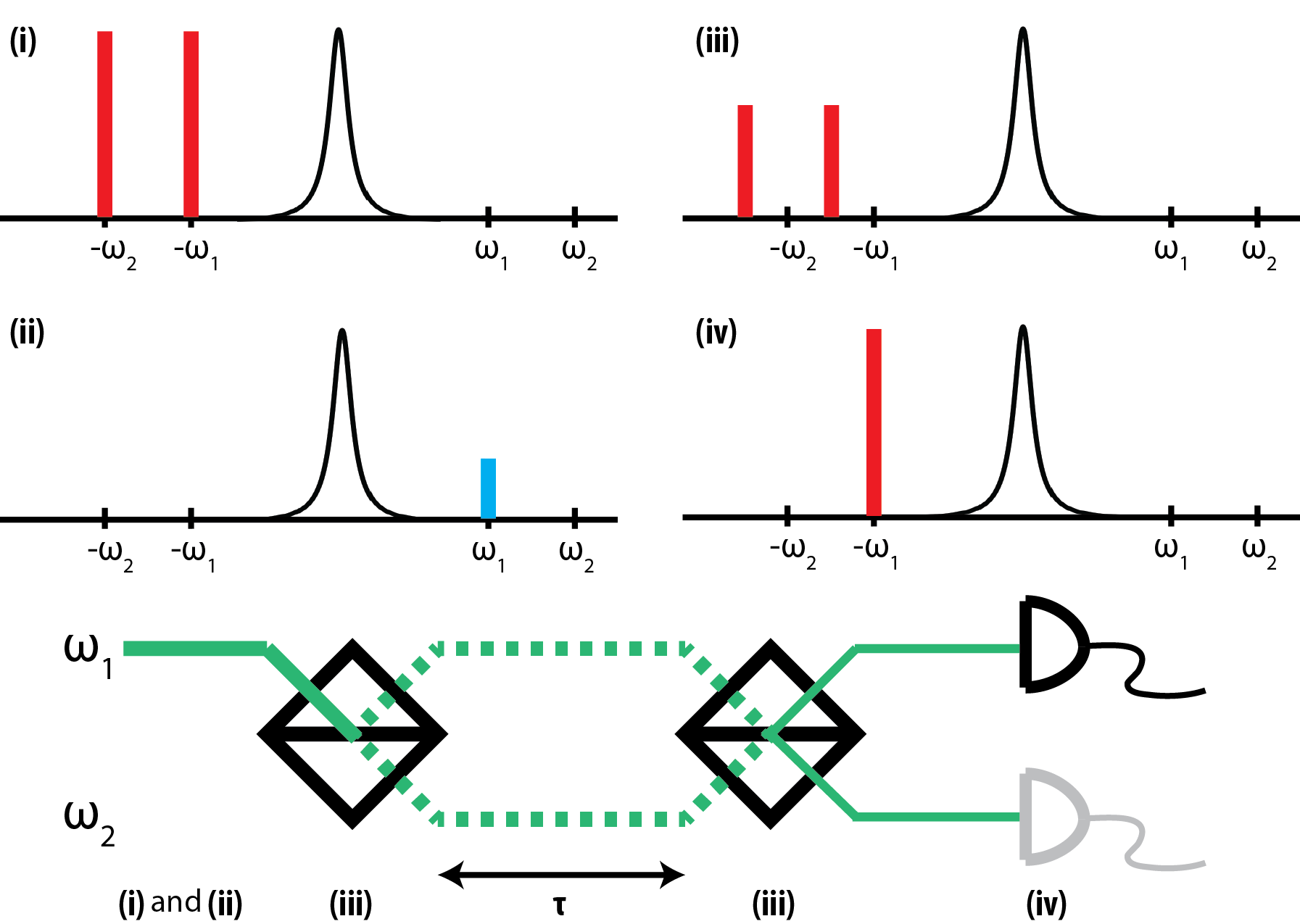}
\caption{This figure shows the four control pulses sent into the optomechanical cavity to execute the experiment. The pulses are: \textbf{(i)} Cooling to the ground state. \textbf{(ii)} Excitation to a coherent state, followed by postselection of the first excited state. \textbf{(iii)} A mechanical-mechanical interaction with $Jt$ = $\pi$/2. \textbf{(iv)} Readout of a resonator. On the bottom, the equivalent optics experiment is shown with the corresponding steps. The greyed out detector is the optional addition of a readout pulse for resonator 2.}
\label{fig:scheme}
\end{figure}

Figure \ref{fig:scheme} illustrates the method we propose to study decoherence. First both mechanical modes must be cooled close to the ground state using standard sideband cooling with two long laser pulses red detuned from the cavity resonance by $\omega_1$ and $\omega_2$ \cite{Marquardt2007,Chan2011,Teufel2011}. Next, we excite resonator 1 to its first excited state using a weak pulse and projection measurement \cite{Galland2014}. We perform a Mach-Zehnder type interference experiment on this initial state. To generate a beam splitter interaction between the mechanical resonators, we apply a two laser pulse, resulting in an entangled state: $\ket{\psi} = \frac{1}{\sqrt{2}}[\ket{1}_1\ket{0}_2 + i\ket{0}_1\ket{1}_2]$. The system now evolves freely for a time $\tau$, possibly decohering during that interval. The frequency difference between the resonators causes the state $\ket{\psi}$ to pick up a phase difference of $(\omega_2-\omega_1)\tau$. A second mechanical-mechanical interaction rotates the system to sin$((\omega_2-\omega_1)\tau/2)\ket{1}_1\ket{0}_2 + $cos$((\omega_2-\omega_1)\tau/2)\ket{0}_1\ket{1}_2$ if the system did not decohere. Finally, a laser pulse red detuned by $\omega_1$ is used to swap the mechanical state of resonator 1 with that of the cavity and read it out with a photodetector. 

We will now examine the steps in more detail, starting with the heralded generation of a single phonon mechanical Fock state \cite{Galland2014}, which has already been used to produce single phonon Fock states with reasonably high fidelity \cite{Riedinger2015,Hong2017}. Here we will review the process briefly, including some of the imperfections in the generated state. A weak pulse of light, blue detuned in frequency by $\omega_1$, is sent into the cavity, creating an effective interaction described by the Hamiltonian: $H_{\textbf{(ii)}} = \hbar\sqrt{n_{cav}}g_1(\hat{a}\hat{b}_1+\hat{a}^\dagger\hat{b}_1^\dagger)$. $n_{cav}$ is the number of photons in the cavity from the laser pulse. This generates an entangled state between the cavity and resonator 1: $\ket{\psi} = 1/\sqrt{2}(\ket{0}_c\ket{0}_1+\sqrt{p}\ket{1}_c\ket{1}_1+p\ket{2}_c\ket{2}_1)$, where $p$ $\ll$ 1 is the excitation probability. The light leaks out of the cavity and passes through a filter to isolate the resonant light from the blue-detuned pulse. By detecting a single photon, the mechanical resonator is projected onto $\ket{1}_1$, a single phonon Fock state. Because of the limited detection efficiency of cavity photons $\eta$, and the dead time of the detector, higher number states will be mistaken as single photons, so the probability $p$ must be kept small to avoid inclusion of these states. Control pulse photons which leak through the filter and detector dark counts will incoherently add in $\ket{0}_1$ to the single phonon Fock state. Using a good filter and superconducting single photon detectors avoids the inclusion of the ground state \cite{Riedinger2015}. Taken together these steps produce, with probability $\eta p$, a heralded single phonon Fock state, and we can proceed to the interference experiment.

Exchange of quantum states is the essence of the interference experiment. In recent years there have been many demonstrations of opto- and electro-mechanically controlled coherent coupling between mechanical resonators \cite{Lin2010,Okamoto2013,Shkarin2014,Noguchi2016,Damskagg2016,Fang2016,Pernpeintner2016}. All of these could be used to create an effective beam splitter interaction between two mechanical resonators. We will use the swapping method proposed by Stamper-Kurn et al. \cite{Buchmann2015}(and experimentally demonstrated in \cite{Weaver2017}), because it is quite general and couples resonators with a large frequency separation, which is important for the individual readout of each resonator. Two pulses of light, red-detuned and separated by $\omega_2-\omega_1$ are sent into the cavity. These pulses each exchange excitations between one mechanical resonator and the cavity mode, resulting in a net swapping interaction with rate $J$ between the two resonators: $H_{\textbf{(iii)}} = \hbar J(\hat{b}_1^\dagger\hat{b}_2+\hat{b}_1\hat{b}_2^\dagger)$. This interaction can be used for both beam splitter interactions in the proposed experiment.

Finally, the readout for the system consists of a pulse of light, red detuned in frequency by $\omega_1$. The readout interaction, $H_{\textbf{(iv)}} = \hbar\sqrt{n_{cav}}g_1(\hat{a}^\dagger\hat{b}_1+\hat{a}\hat{b}_1^\dagger)$, exchanges excitations of resonator 1 with photons on resonance in the cavity. The anti-Stokes photons from the cavity are filtered and sent to a superconducting single photon detector to determine the phonon occupation of resonator 1 with a collection efficiency of $\eta$. Because of the difference in frequency of the two resonators, the measured phonon occupation of resonator 1 after the second mechanical-mechanical interaction oscillates as a function of the delay time $\tau$ at the frequency $\omega_2-\omega_1$. However, if decoherence occurs during free evolution, the visibility of the oscillations will decrease. These features in the readout enable a simultaneous comparison of the coherent evolution, decoherence and thermalization of the system.

\section{Expected Results}

\begin{figure}
\includegraphics[width=8.6cm]{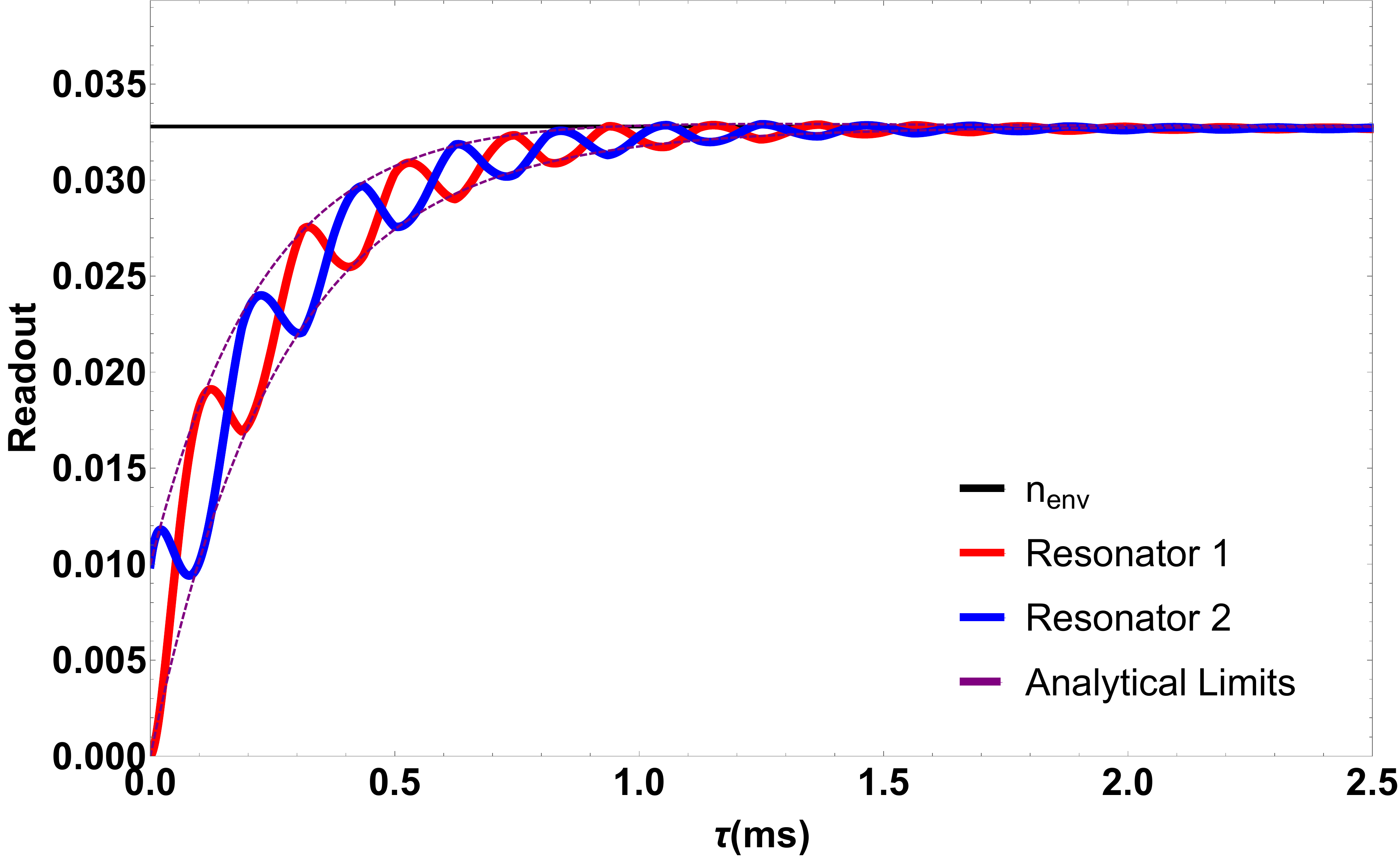}
\caption{Expected results of a decoherence measurement with two entangled resonators in which one interacts with a thermal environment. The red (blue) indicates the readout of resonator 1 (2). Dotted lines are the limits set by the analytical model. Three effects are visible: coherent oscillations due to the frequency difference between the resonators, a decay of that coherence due to environmentally induced decoherence, and thermalization with the environment. The parameters for this plot are: $\omega_1$ = 2 GHz, $\Delta\omega$ = 30 kHz, $\gamma$ = 2 kHz, $T_{env}$ = 0.1K, and $\eta$ = 0.01.}
\label{fig:results}
\end{figure}

First we model the experiment analytically. We assume that in step \textbf{(ii)} of Figure \ref{fig:scheme} a perfect entangled state is generated, but that the off-diagonal elements of the density matrix decay exponentially with a decoherence time $\tau_d$. The environment heats resonator 2, adding incoherently to the mechanical state. As an approximation, we assume that the state thermalizes from its average initial value of 1/2 to the thermal occupation of the environment, $n_{env}$. The average readout, $R$ on the SSPD in step \textbf{(iv)} after many trials is the sum of the two effects:

\begin{subequations}
\label{allequations}
\begin{eqnarray}
\braket{n_{dec}}_2 &=& \frac{1}{2}-\frac{\cos\left[(\omega_2-\omega_1)\tau\right]e^{-\tau/\tau_d}}{2} \\
\braket{n_{th}}_2 &=& \left(n_{env}-\frac{1}{2}\right)\left(1-e^{-\tau/\tau_{th}}\right) \\
R &=& \eta\left(\braket{n_{dec}}_2 + \braket{n_{th}}_2\right)
\end{eqnarray}
\label{eq:readout} 
\end{subequations}
$n_{env}$=$k_BT_{env}/\hbar\omega_2$ is the thermal occupation of the environment at temperature $T_{env}$ and $\tau_{th}$ is the thermalization time constant. Three key features are visible in the readout signal: an oscillation at $\omega_2$ -$\omega_1$ which is evidence of coherence, an exponential decay of the coherent signal and an exponential increase in the phonon number as the system thermalizes. 

\begin{figure*}
\includegraphics[width=17cm]{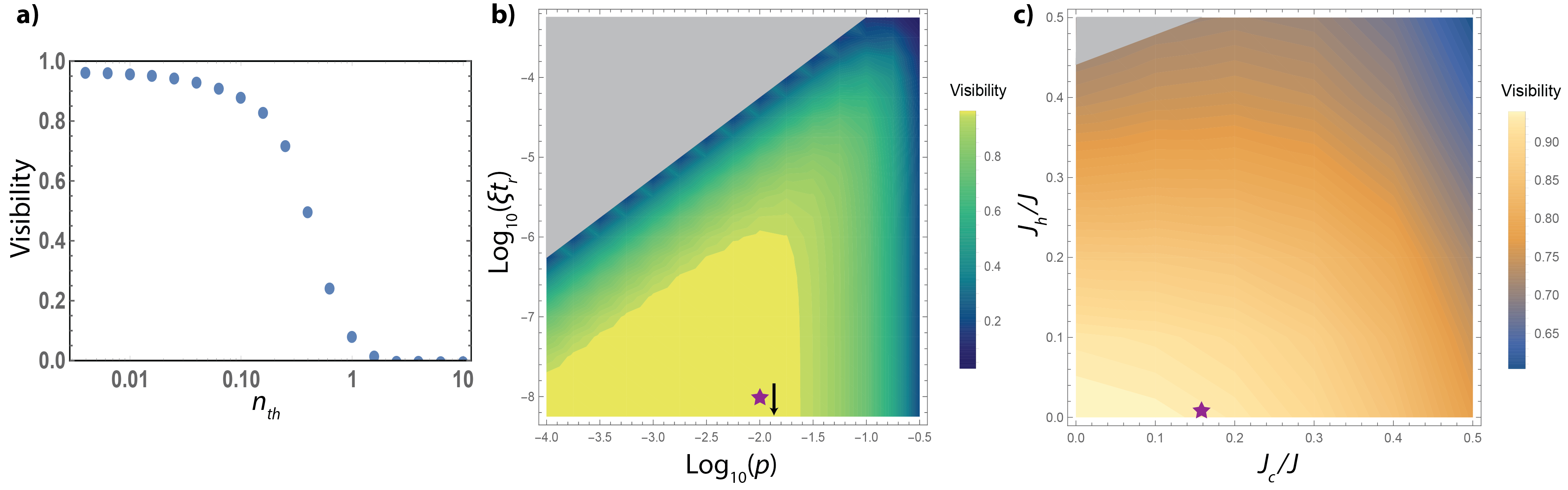}
\caption{The effects of several experimental imperfections on the resulting interference experiment, using the initial visibility as a metric. \textbf{a)} The two resonators are only cooled to a phonon occupancy of $n_{th}$ in step \textbf{(i)}. \textbf{b)} The detector has a dark count probability of $\xi t_r$ for different probabilities of excitation, $p$ in step \textbf{(ii)}. \textbf{c)} Step \textbf{(iii)} also induces an optical cooling rate $J_c$ and an optical heating rate $J_h$ in addition to the mechanical-mechanical coupling $J$. The greyed out regions indicate regimes in which the dominant behavior is not the desired entangled state. The purple stars indicate parameters already achieved in experiments: \textbf{b)}\cite{Hong2017} and \textbf{c)} \cite{Weaver2017}. The unvaried parameters for these plot are: $n_{th}$=0.01, $ p$=0.01, $ \xi t_r$=$10^{-6}$, $ \eta$=0.01, and $ J_c$=$J_h$=0. }
\label{fig:imperfections}
\end{figure*}

We verify Equation \ref{eq:readout} by performing a numerical simulation of the interaction between a mechanical resonator and its environment in the quantum master equation formalism. We assume that one resonator, the test mass resonator, has a much greater interaction rate $\gamma$ with the environment, dominating the decoherence effects. Environmentally induced decoherence can be modeled as an interaction with a bath of harmonic oscillators, leading to the following master equation \cite{Caldeira1983,Zurek2003}:

\begin{equation}
\dot{\rho} = \frac{i}{\hbar}\left[\rho,\hat{H}_0\right] - \frac{D}{\hbar^2}\left[\hat{x},\left[\hat{x},\rho\right]\right] - \frac{i\gamma}{\hbar}\left[\hat{x},\{\hat{p},\rho\}\right]
\label{eq:master}
\end{equation}
$\hat{x}$ and $\hat{p}$ are the position and momentum operators for resonator 2, and $D$=$2m\gamma k_BT_{env}$ is the phonon diffusion constant. The numerical results are shown in Figure \ref{fig:results}, and have excellent agreement with Equation \ref{eq:readout}.

We now discuss the experimental feasibility of this scheme with currently available technologies. We numerically simulate density matrices with the phonon states of each resonator as basis states. (Details in Appendix \ref{appendix}.) The initial visibility of the oscillations between the two resonators is a direct measure of the entanglement generation, and the decay of the visibility is the essential result of the experiment. Although the limit would depend on the exact experimental implementation, we estimate that the experiment would likely require an initial visibility greater than 10\%. First we consider imperfections in step \textbf{(i)}, cooling to the ground state. Figure \ref{fig:imperfections}a shows the visibility achieved with a nonzero thermal phonon occupation. This occupation must be below about 0.7 for the experiment to be feasible.

Next we consider step \textbf{(ii)}, the postselection of a single phonon state. By changing the pulse strength, the probability $p$ of an excitation can be adjusted. Dark counts on the single photon counter during the postselection will skew the produced state. Figure \ref{fig:imperfections}b shows the visibility as a function of $p$ and dark count rate. There is a large region of parameter space with good visibility, and experiments are already well within this region (purple star) \cite{Hong2017}. 

\begin{figure}
\includegraphics[width=8.6cm]{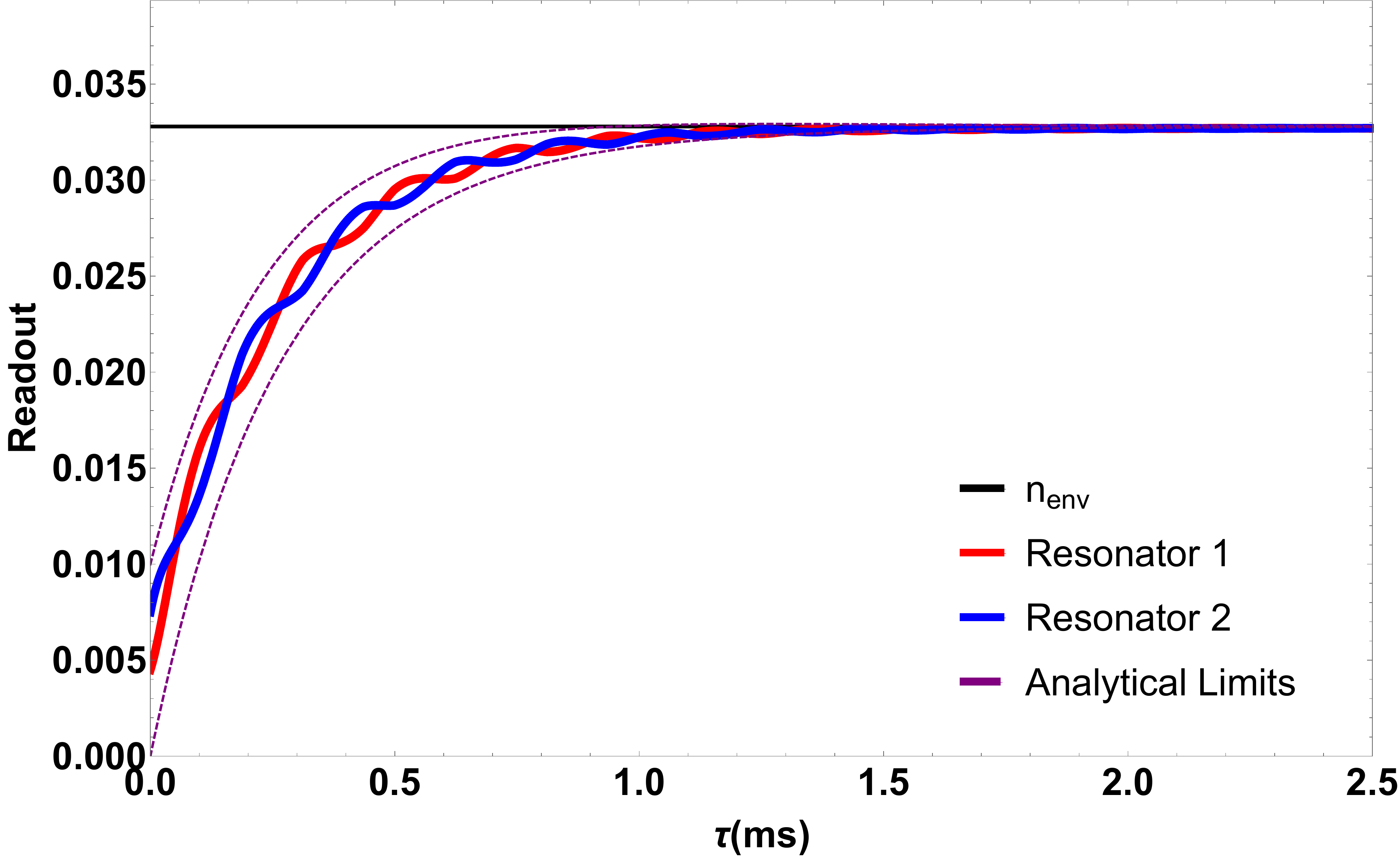}
\caption{Expected results of a decoherence measurement with imperfections present and an initial visibility of 30\%. The parameters for this plot are the same as for Figure \ref{fig:results} with additional imperfections: $n_{th}$ = 0.4, $p$ = 0.1, $J_c$ = $J_h$ = 0. Despite the limited initial visibility all three effects are still visible: coherent oscillations due to the frequency difference between the resonators, a decay of that coherence due to environmentally induced decoherence, and thermalization with the environment. We estimate that the experimental limit on the initial visibility is around 10\%.}
\label{fig:resultsimperfections}
\end{figure}

Finally, in step \textbf{(iii)}, the optomechanical beam splitter nominally only causes an interaction between the two mechanical resonators. However, the beams used to produce the interaction also have heating and cooling effects. In Figure \ref{fig:imperfections}c the visibility as a function of cooling rate, $J_c$ and heating rate $J_h$ are shown. Again, experimental demonstrations of this type of beam splitter interaction are already sufficient to produce an interference experiment \cite{Weaver2017}. In Figure \ref{fig:resultsimperfections} we show numerical simulations of decoherence and thermalization that include experimental imperfections and an initial visibility of 30\%. All of the qualitative features of Figure \ref{fig:results} are still easily discernable, indicating that the experiment should be feasible with these or even slightly worse parameters. There is a large area of experimentally achievable parameter space in all dimensions with visibility greater than 10\%. 

\section{Timing Considerations}

A number of experimental factors such as timing also play a critical role in the feasibility of the experiment. The probability of a successful postselection is $\eta p$, and given this successful postselection the probability of measuring the result on the detector is $\eta$. Therefore, the experiment must be run 1/$\eta^2 p \sim 10^6$ times to expect a single detection event. For many experimental implementations this is impossible, because it would take years to build up enough detection events. However, if there is no heralding of a single photon in step \textbf{(ii)}, there is no reason to continue the experiment. If we only continue to step \textbf{(iii)} after a successful postselection the time $T$ required is:

\begin{equation}
T = n_{a}n_{p}\left(\frac{t_{12}(1-\eta p)}{\eta^2p} +  \frac{t_{tot}\eta p}{\eta}\right) \\ \approx n_{a}n_{p}\frac{t_{12}}{\eta^2p}
\end{equation}
$t_{12}$ and $t_{tot}$ are the time required for step \textbf{(i)} and \textbf{(ii)} and for the total experiment respectively, and $n_a$ and $n_p$ are the number of averages and the number of points. In general, step \textbf{(iii)} and $\tau$ should dominate the experiment time, so this would drastically reduce the total experiment time. For a high frequency resonator with $\sim$GHz frequency, reasonable parameters might be: $n_a$ = 1000, $n_p$ = 30, $\eta$ = 0.01, $p$ = 0.01 and $t_{12}$ = 1$\upmu$s, leading to an experiment time of about 8 hours. For lower frequency resonators, $t_{12}$ might be closer to 100 $\upmu$s, leading to an experiment time of about 35 days. The number of averages needed depends inversely on $\eta$, so $T\sim1/\eta^3$, and the experiment can be drastically sped up by increasing $\eta$.

Many experiments which are proposed for testing novel decoherence mechanisms are in the lower frequency range. These experiments have the difficulty that their thermal environment contains more thermal quanta. In order to measure the full thermalization in addition to the decoherence, we must be able to count $\eta n_{env}$ photons. If an SSPD has a relatively short dead time ($\sim$100 ns) compared to the leakage time from the cavity and filter ($\sim$50 $\upmu$s) it may be possible to observe more than one photon. In general, however, the experiment should be constrained to $\eta \ll 1/n_{env}$. For low frequency resonators $\eta$ may need to be artificially lowered. If this is the case, we suggest different detectors for step \textbf{(ii)} and step \textbf{(iv)} with different optical paths. If step \textbf{(ii)} has high efficiency $\eta_1$ and step \textbf{(iv)} has low efficiency $\eta_2$ the experiment time only slows down to $T\approx n_{a}n_{p}t_{12}/\eta_1\eta_2p$ and it is possible to count higher phonon numbers with a reasonable increase in experiment time. 

\section{Experimental Implementations}

This scheme can be performed with any two mechanical resonators coupled to an optical cavity. Here we will discuss three potential experimental setups, with an emphasis on using the technique to access decoherence information in large mass systems. One possible system is a Fabry-P\'{e}rot cavity with two trampoline resonators: one with a distributed bragg reflector (DBR) and one without. This system has already been constructed \cite{Weaver2017}. The two resonators have frequencies in the hundreds of kHz range, a mass of 40 ng and 150 ng and a single photon cooperativity 0.0002 and 0.0001 respectively. The authors suggest methods for lowering optical and mechanical damping, which would improve the single photon cooperativity to 0.2 and 0.01. The scheme presented here enables single phonon control of the massive DBR device despite its relatively small single photon cooperativity.

Another possible system would be a membrane in the middle at one end of a Fabry-P\'{e}rot cavity and a cloud of atoms trapped in the harmonic potential of the standing wave in the cavity at the other end. The optomechanical coupling enables the direct coupling between the $\sim$zg cloud of atoms and the $\sim$100 ng membrane. Clouds of atoms and membranes have already been coupled between different cavities \cite{Jockel2015,Zhong2017}, and this scheme could be modified to use that interaction for step \textbf{(iii)}. One could also imagine making a cavity with a bulk acoustic wave resonator coupled to a small high frequency membrane. These modes can have exceptionally high Q-factors and large mode mass \cite{Goryachev2012}. 

\section{Discussion}

There are a number of distinct advantages of the method proposed here. First, the readout of phonon occupation naturally lends itself to studying thermalization and decoherence together in the same system and on the same time scale. This has never been observed before in mechanical resonators. A thorough understanding of the mechanics of thermalization and decoherence is necessary in order to verify that unknown faster decoherence processes can be attributed to new physics. Second, this experiment can easily be compartmentalized into the four constituent steps, and each one tested individually. This would make it easier to build up to the final experiment with confidence in the results. In particular, one could obtain interference results from two resonators in a classical state, so it is essential to demonstrate that the procedure is performed with a single phonon. Finally, this scheme can use mechanical resonators with different frequencies and masses, so that large systems with relatively small optomechanical coupling rates can be studied.

\section{Conclusion}

We have proposed a scheme to entangle two mechanical resonators with a shared single phonon. Using interferometry and phonon counting we could simultaneously measure decoherence and thermalization of a macroscopic mechanical mode. The methods proposed are quite general, and can be applied to any sideband resolved two mode opto- or electro-mechanical system. Furthermore, the scheme is resilient to experimental imperfections in its constituent steps. This technique could greatly expand our understanding of the quantum to classical transition in mechanical systems.

\section{Acknowledgements}

The authors would like to thank F. Buters, H. Eerkens, S. de Man, V. Fedoseev and S. Sonar for helpful discussions. This work is part of the research program of the Foundation for Fundamental Research (FOM) and of the NWO VICI research program, which are both part of The Netherlands Organisation for Scientific Research (NWO). This work is also supported by the National Science Foundation Grant Number PHY-1212483.

\appendix

\section{Numerical Methods \label{appendix}}

In the main text we investigate two main problems. The first is the interaction of a mechanical entangled state with the bath of one resonator. We use a numerical differential equation solver to solve the Master Equation (Equation \ref{eq:master}) with density matrices. After some algebraic manipulation, this can be rewritten as a set of differential equations:

\begin{figure}
\includegraphics[width=8.6cm]{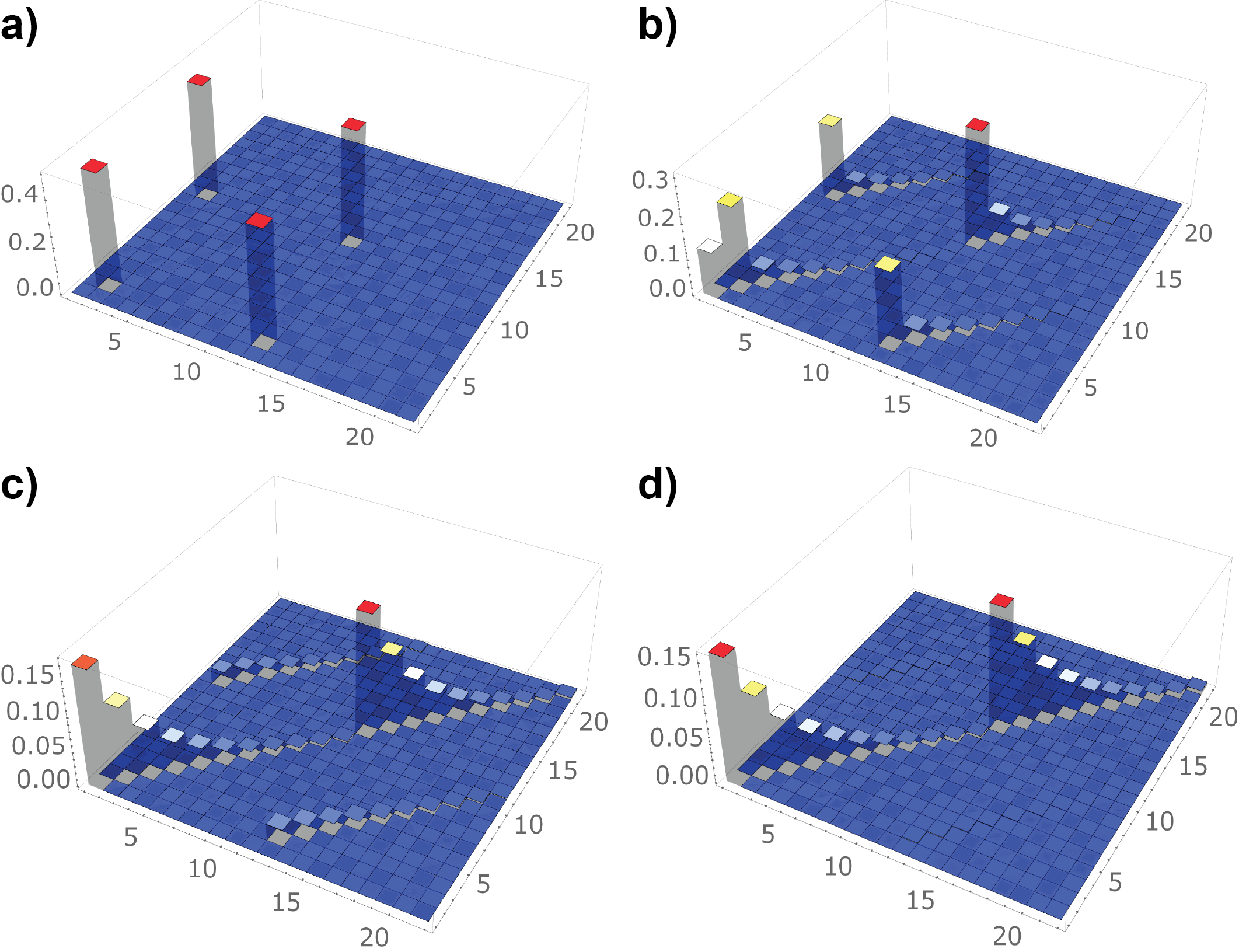}
\caption{Density matrix representation of decoherence and thermalization. Each matrix is plotted during step \textbf{(iii)} after a delay time $\tau$ of \textbf{a)} 0 ms \textbf{b)} 190 ms \textbf{c)} 950 ms \textbf{d)} 3.8 s. The states labeled 1 to 22 in the figure correspond to the basis states $\{00,01,...,09,010,10,11,...,19,110\}$. The relevant parameters are $\omega_2$ = 10 GHz, $\gamma$ = 1 Hz and $T_{env}$ = 0.2 K.}
\label{fig:densitymatrix}
\end{figure}

\begin{eqnarray}
\rho &=& \sum_{p,q,r,s=0}^\infty a_{pqrs}(t)\ket{pr} \bra{qs} \\
\left[\hat{x},\left[\hat{x},\ket{r}_2\bra{s}_2\right]\right] &=&\sum_{k,l} \Gamma_{klrs} \ket{k}_2\bra{l}_2 \\
\left[\hat{x},\left\{\hat{p},\ket{r}_2\bra{s}_2\right\}\right] &=&\sum_{k,l} \Phi_{klrs} \ket{k}_2\bra{l}_2 \\
\dot{a}_{pqrs}(t) &=& -i\left(\omega_1(p-q)+\omega_2(r-s)\right)a_{pqrs}(t) \nonumber \\ &-& \frac{D}{\hbar^2}\sum_{k,l=0}^\infty \Gamma_{rskl}a_{pqkl}(t) \nonumber
\\ &-& \frac{i\gamma}{\hbar}\sum_{k,l=0}^\infty \Phi_{rskl}a_{pqkl}(t)
\end{eqnarray}
The commutation relationships in the equations lead to a number of overlap integrals between number states, which can be evaluated and plugged in to create numerically solvable equations. To solve for the dynamics of this system we use a density matrix with basis states \{00,01,...0$n$,10,11,...1$n$\} where $n$ is a number much larger than $n_{env}$. Figure \ref{fig:densitymatrix} shows the results of the simulations for $n$=10 at four different times before the second swapping pulse. Two main effects are observable in the evolution of the density matrix. First, the population of the density matrix spreads out along the diagonal of each of the four quadrants. Second, the non-diagonal matrix elements decay away. These effects match with the expected behavior for thermalization and decoherence.  

We also need to simulate a mechanical-mechanical $\pi$/2 pulse. Because it is equivalent to a beam splitter the effect on the two modes is the same. Here we expand the density matrix to have basis states \{00,01,...0$n$,10,11,...1$n$,$n$0,$n$1,...$nn$\}. The beam splitter interaction conserves energy, so it can represented as a $n^2$x$n^2$ transformation matrix, which recombines the elements of common phonon number. The transformation matrix $S_{BS}$ for the three lowest energy levels with basis states \{00,01,10,02,11,20\} is:

\begin{equation}
S_{BS} = \left(\begin{tabular}{cccccc}
1 & 0 & 0 & 0 & 0 & 0 \\
0 & 1/$\sqrt{2}$ & -1/$\sqrt{2}$ & 0 & 0 & 0\\
0 & 1/$\sqrt{2}$ & 1/$\sqrt{2}$ & 0 & 0 & 0 \\
0 & 0 & 0 & 1/2 & -1/$\sqrt{2}$ & 1/2 \\
0 & 0 & 0 & -1/2 & 0 & 1/2 \\
0 & 0 & 0 & 1/2 & 1/$\sqrt{2}$ & 1/2
\end{tabular}\right)
\end{equation}
After the beam splitter interaction the density matrix $\rho'$ is $S_{BS}^T\rho S_{BS}$. The combination of these two techniques lets us fully model how the ideal state interacts with its thermal environment.

The other problem we investigate is how various experimental imperfections can impact the initial visibility of the experiment. For this we use density matrices with basis states going up to $n$=3. To model imperfect cooling in step \textbf{(i)} we start with a thermal state of both resonators. The modeling of step \textbf{(ii)} is a little more complex. A successful postselection means that 1 phonon has been added to resonator 1. However, with probability $p$, the phonon occupation should be incremented by 2, and with probability $p^2$ by 3, and so on. Conversely, if there is a dark count or leaked pulse photon (probability $\xi t_r$) the phonon occupation should remain the same. Finally, we implement the beam splitter, step (iii), in the same way as above. We add in an additional cooling pulse with a probability $J_c/J$ of removing a phonon from one of the resonators and a heating pulse with a probability $J_h/J$ of adding a phonon to a resonator. The cooling matrix transformation $S_c$ with basis states \{00,01,02,10,11,12,20,21,22\} is:

\begin{equation}
S_c = \left(1-\frac{J_c}{J}\right)I+\frac{J_c}{J}\left(\begin{tabular}{ccccccccc}
0 & 1 & 0 & 1 & 0 & 0 & 0 & 0 & 0\\
0 & 0 & $\sqrt{2}$ & 0 & 1 & 0 & 0 & 0 & 0\\
0 & 0 & 0 & 0 & 0 & 1 & 0 & 0 & 0 \\
0 & 0 & 0 & 0 & 1 & 0 & $\sqrt{2}$ & 0 & 0 \\
0 & 0 & 0 & 0 & 0 & $\sqrt{2}$ & 0 & $\sqrt{2}$ & 0 \\
0 & 0 & 0 & 0 & 0 & 0 & 0 & 0 & $\sqrt{2}$ \\
0 & 0 & 0 & 0 & 0 & 0 & 0 & 1 & 0 \\
0 & 0 & 0 & 0 & 0 & 0 & 0 & 0 & $\sqrt{2}$ \\
0 & 0 & 0 & 0 & 0 & 0 & 0 & 0 & 0
\end{tabular}\right)
\end{equation}
The heating matrix transformation is $S_c^T$. All of these imperfections are combined to determine their impact on the proposed experiment.

\section{Additional Experimental Considerations \label{additionalexp}}

The first additional consideration relates to the pulses used in the experiment. It is possible to perform the experiment with simple square-shaped pulses. However, it is more efficient to use an exponentially shaped pulse, resulting in a more even interaction time \cite{Hofer2011}. We suggest using pulses of that shape, as is performed in \cite{Hong2017}. In particular, it is crucial that the area under the readout pulse: $\int_0^\infty n_{cav}(t)g_1 dt$ is $\pi/2$ to fully readout the phonon occupation of resonator 1.

\begin{figure}
\includegraphics[width=8.5cm]{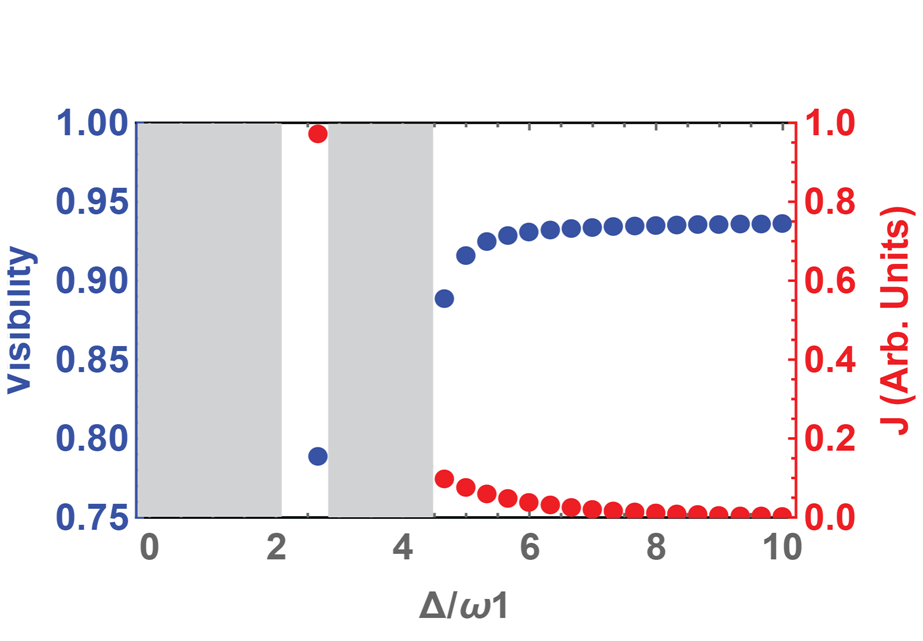}
\caption{Visibility as a function of detuning. Greyed out regions have too high $J_c$ or $J_h$ to run the experiment. The value of $J$ depends on the exact experimental parameters, so it is normalized to the highest value. Parameters are $\omega_2/\omega_1$ = 2, $\omega_1/\kappa$ = 10, $n_{th}$=0.01, $ p$=0.01, $ \xi t_r$=$10^{-6}$ and $ \eta$=0.01.}
\label{fig:detuningsweep}
\end{figure}

We also consider the most effective detuning of the two laser beams for performing a $\pi$/2 pulse. The two laser tone exchange method relies on exchanging the state of each mechanical resonator with that of the cavity. This is fastest if the two laser beams are red detuned to $\omega_1$ and $\omega_2$. However, at this detuning quantum information leaks out of the cavity, leading to large values of $J_c$ and $J_h$. In Figure \ref{fig:detuningsweep} we examine the effects of the average detuning $\Delta$ of these two laser beams. Ideally the two beams should be quite far detuned from the cavity, but there is a tradeoff between efficient exchange and the exchange rate, $J$, shown in red \cite{Weaver2017}. The best detuning depends on experimental parameters such as sideband resolution and frequency of the resonators.

\bibliographystyle{apsrev}
\bibliography{phononinterferometry}

\end{document}